\documentclass[aps,prb, 10pt, twocolumn]{revtex4-2}
\usepackage{amsmath}
\usepackage{amsfonts}
\usepackage{amssymb}
\usepackage{graphicx}
\usepackage{bm}
\usepackage{hyperref}
\usepackage{xcolor}

\begin{document}

\title{Exploring exotic configurations with anomalous features  using deep learning:\\ Application of classical and quantum-classical hybrid anomaly detection}

\author{Kumar J. B. Ghosh$^{1}$}
\email{jb.ghosh@outlook.com}

\author{Sumit Ghosh$^{2,3}$}
\email{s.ghosh@fz-juelich.de}

\affiliation{$^1$E.ON Digital Technology GmbH,  45131, Essen, Germany\\
$^2$~Institute of Physics, Johannes Gutenberg-University Mainz, 55128 Mainz, Germany.\\
$^3$~Institute of Advance Simulations, Forschungszentrum J{\"u}lich GmbH, 52428 J{\"u}lich, Germany}


\begin{abstract}
In this article we present the application of classical and quantum-classical hybrid anomaly detection schemes to explore exotic configuration with anomalous features. We consider the Anderson model as a prototype where we define two types of anomalies - a high conductance in presence of strong impurity and low conductance in presence of weak impurity - as a function of random impurity distribution. Such anomalous outcome constitutes an imperceptible fraction of the data set and is not a part of the training process. These exotic configurations, which can be a source of rich new physics, usually remain elusive to conventional classification or regression methods and can be tracked only with a suitable anomaly detection scheme. We also present a systematic study of the performance of the classical and the quantum-classical hybrid anomaly detection method and show that the inclusion of a quantum circuit significantly enhances the performance of anomaly detection which we quantify with suitable performance metrics. Our approach is quite generic in nature and can be used for any system that relies on a large number of parameters to find their new configurations which can hold exotic new features.

\end{abstract}

\maketitle


\section{introduction}

In recent years machine learning has become an integral part of different branches of condensed matter physics. It has shown impeccable performance is dealing with problem with large degrees of freedom where extracting an effective model is practically impossible. It has been adopted as a viable alternative for exploring electronic properties \cite{Chandrasekaran2019, Sheverdin2020, Westermayr2021, Kulik2022}  as well as transport properties \cite{Lopez-Bezanilla2014, Li2020a, Kotthoff2021, Ghosh2023}. Its inherent ability to deal with a high level non-linearity makes it quite successful in highly non-trivial physical problem such as predicting different phases of matter \cite{Carrasquilla2017, Tibaldi2023} and their topological characterisation \cite{Claussen2020}. In addition to playing a crucial role in discovering new materials as well as mapping their quantum features \cite{Torlai2018}, this has been instrumental in designing new experiments to unravel their quantum nature \cite{Krenn2016}. Although such automatization makes it possible to scan through a huge configuration space, it also has a risk of missing exotic configurations containing significantly new physics. Occurrence of such configurations are statistically insignificant and can be easily overlooked in a learning process. Identifying these rare configurations therefore can hold key to discovering new physics. 

In this article, we present a new paradigm, namely \textit{anomaly detection} \cite{Omar2013MachineLT, valdes2016anomaly, cui2018anomaly} which is particularly suitable for detecting such special configurations. The main advantage of anomaly detection with respect to conventional classification schemes is that here one doesn't need the a priory knowledge of the data points that are uncharacteristic for a specific data set, namely \textit{anomaly}. The training is done with normal data. The anomalies are heterogeneous and remain unknown until their occurrence. For example consider the ECG of regular heart beat which shows a periodic pattern. An anomaly detection algorithm trained with the normal heart beat can identify the irregularities which has not been observed before and can predict signatures of heart problems \cite{Lima2019, Colt2021}. Due to the rarity of anomalous events, anomaly-detection data sets are heavily imbalanced. It is, therefore, a highly complex task to formally describe an anomaly \cite{deep_anomaly}.

In this work, we demonstrate how anomaly detection can be exploited to reveal subtle features of a condensed matter system which can remain hidden from any conventional regression or classification scheme. We consider the Anderson model where the distribution of the random impurity constitute the input parameter space. The output is the conductance of the system which falls down significantly for strong impurity strength. However for certain distribution the system might achieve a significantly larger transmission which we mark as \textit{anomaly}. Such occurrence is statistically insignificant and therefore almost impossible to anticipate beforehand. However a configuration which can provide large conductance in presence of strong impurity can be quite useful in device design. For example, disorder can enhance damping like spin-orbit torque, which is responsible for electrical switching of magnetisation \cite{Ghosh2017}, or it can enhance the superconducting nature \cite{Neverov2022} as well. On the other hand, for a weak impurity strength, when the system is expected to have a high transmission, the anomaly is defined as the configuration which suppresses the conductance significantly. Such anomalies can pose a hurdle in quantum optimisation even in presence of weak impurity \cite{Altshuler2010}. The main objective of the present work is to systematically identify these anomalies with a machine learning algorithm which can be utilised to understand the nature of such unusual configurations. 

In this paper we demonstrate the application of both classical \cite{deep_anomaly} and and quantum-classical hybrid anomaly detection scheme \cite{Romero_2017, qGAN4anomaly, anogan, sakhnenko2021hybrid} for physical problems that manifests anomalous behaviour as a complex function of large number of variables. Taking the Anderson model as a prototype,  we systematically show that how a anomaly detection scheme can identify the outliers without any prior knowledge of their existence. We consider three methods namely isolation forest \cite{isolation_forest}, auto encoder \cite{baldi2012autoencoders, makhzani2015adversarial, vincent2010stacked} and hybrid quantum-classical auto encoder \cite{sakhnenko2021hybrid} and compare their performances in terms of suitable performance metrics. Our analysis shows that the quantum anomaly detection schemes performs better compared to their classical counterpart due to their inherent ability to deal with the complex feature mapping in the latent dimension. Note that, although the present work is focused on anomalous behaviour of conductance due to impurity scattering, the framework is applicable for detecting anomaly in any physical observable as a function of arbitrarily large number of parameters and therefore would play an instrumental role in discovering new exotic configurations for different physical systems.  


\section{Model and method \label{model}}

\begin{figure}[h!]
\centering
\includegraphics[width=\linewidth]{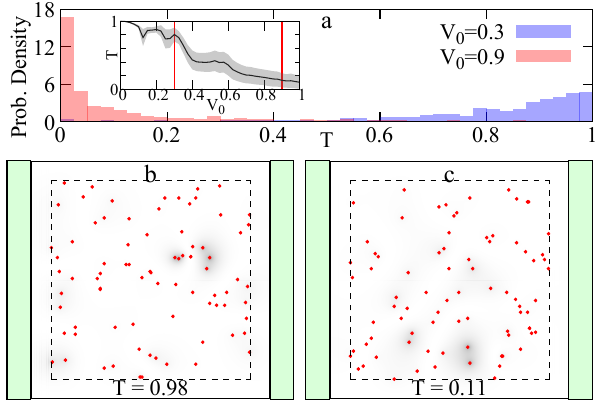}
\caption{Data distribution and schematic of the device configuration. (a) Distribution of conductance for $V_0$=0.9 (red) and $V_0$=0.3 (blue). Inset shows the variation of conductance (solid black line, grey region shows the rms deviation) with vertical red line denoting $V_0$=0.3.0.9. (b) and (c) show two configurations for $V_0$=0.9 with local dos (in gray scale) and conductance (in legend). The green regions show the electrodes. The red dots denotes the impurities which are confined within the central region marked by black dashed line.}
\label{fig:1}
\end{figure}

For our study we consider the Anderson model given by
\begin{eqnarray}
H = \epsilon \sum_{i} c_i^{\dagger} c_i + t \sum_{\langle i, j \rangle} c_i^{\dagger} c_j + \sum_{i} V_i c_i^{\dagger} c_i
\end{eqnarray}
where $c^\dagger,c$ is the creation/annihilation operator. $t$ is the hopping parameter which we choose to be -1. $\epsilon$ is the onsite energy which we choose as $-4t$. $V_i$ in the onsite random potential. For this study we consider a $240 \times 240$ scattering region and use it in a two terminal device configuration (Fig.\,\ref{fig:1}). We choose total 80 impurities  with same strength distributed within a $200 \times 200$ region in the centre. We assign a constant negative value $-V_0$ for all the impurities. The Fermi level is kept at $0.0005t$ which gives a conductance of 1 in clean limit. The zero bias conductance of the system is given by 
\begin{equation}
T = Tr\left[\Gamma_1 G^R \Gamma_2 G^A\right] ,
\end{equation}
where $G^{R,A}=\left[E-H-\Sigma^{R,A}_1-\Sigma^{R,A}_2\right]^{-1}$ is the retarded/advanced Green's function of the scattering region.  $\Gamma_{1,2} = i\left[\Sigma_{1,2}^R - \Sigma_{1,2}^A\right]$ where $\Sigma_{1,2}^{R,A}$ is the retarded/advanced self energy of the left/right electrode. For our calculation we use tight binding code KWANT \cite{Groth2014} which uses scattering wave function formalism to obtain these quantities. The local density of states can be obtained directly from the scattering wave function.

From Fig.\,\ref{fig:1}a one can see that for a strong impurity strength, the conductance of the system is more likely to be close to zero. However, some configurations can also give rise to a high value of conductance although the probability of such outcome is quite small. For strong impurity we label such outcome as \textit{anomaly}. Similar behaviour can be observed with a weak impurity. The \textit{anomaly} in this case is a configuration that can completely suppress the current flow resulting a insulating behaviour. From Fig.\,\ref{fig:1} one can see that the impurity configurations corresponding to a high (anomaly) and low (normal) value of conductance don't have any characteristic difference. It is therefore impossible to detect such anomalous behaviour with any conventional method only from the knowledge of the distribution. In the following,  we are going to show how an anomaly detection algorithm can detect such anomalies without any a priory knowledge of such outcome.


\subsection{Classical anomaly detection}
\label{subsection:anomalies_in_autoencoder}

Here, we summarise two different classical machine learning methods we use for anomaly detection. The first method is called isolation forest (IF) \cite{isolation_forest}, which is an unsupervised anomaly detection algorithm that uses a random forest algorithm under the hood to detect outliers in the data set. The algorithm tries to isolate the data points using decision trees such that each observation gets isolated from the others.

\begin{figure}[htb]
\centering
\includegraphics[width=\linewidth]{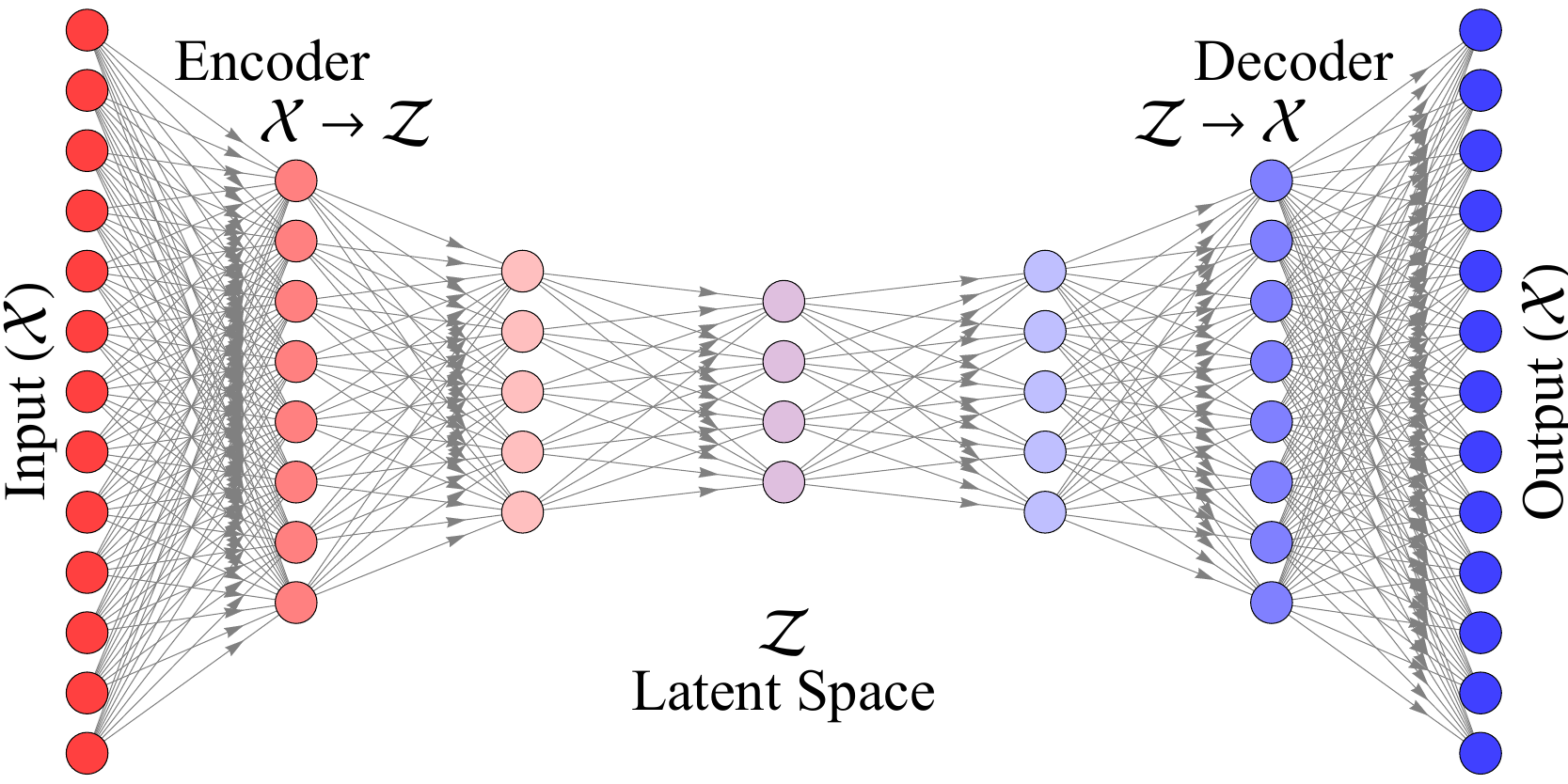}
\caption{Schematic representation of an autoencoder. The input data is compressed by the encoder and the decoder expands the compress data to its original size. The intermediate space with compressed dimension is called the latent space.}
\label{fig:auto-encoder}
\end{figure}

The second method is called \textit{autoencoder} (AE) \cite{baldi2012autoencoders, makhzani2015adversarial, vincent2010stacked}, which is a deep neural network architecture (Fig.\ref{fig:auto-encoder}). It aims to learn a compressed representation for an input through minimizing its reconstruction error \cite{asperti2020balancing, sabokrou2016video}. It consists of two parts - an \textit{encoder} ($e$) and a \textit{decoder} ($d$). The encoder learns a non-linear transformation  $e : \mathcal{X} \to \mathcal{Z}$, that projects the data from the original high-dimensional input space $\mathcal{X} \equiv \lbrace x \rbrace$ to a lower-dimensional latent space $Z \equiv \lbrace z \rbrace$. For our study we consider a latent space with 4 nodes. A decoder learns a non-linear transformation $d : \mathcal{Z} \to \mathcal{X}$ that projects the latent vectors $z=e(x)$ back into the original high-dimensional input space  $\mathcal{X}$. This transforms the latent vector $z=e(x)$ and reconstruct the original input data as $\hat{x}= d(z)=d\left(e(x)\right)$,  where $\hat{x}$ is the output corresponding to an input $x$. One can obtain a more robust decoding of latent vectors with a \textit{variational autoencoder} (VAE) \cite{kingma2013auto}, which is a neural network that unify variational inference approaches with autoencoders. For our study we focus only on IF and AE.

\subsection{Hybrid Quantum-Classical Autoencoder (HAE)} \label{section:hybrid_autoencoder}

\begin{figure}[h!]
\centering
\includegraphics[width=\linewidth]{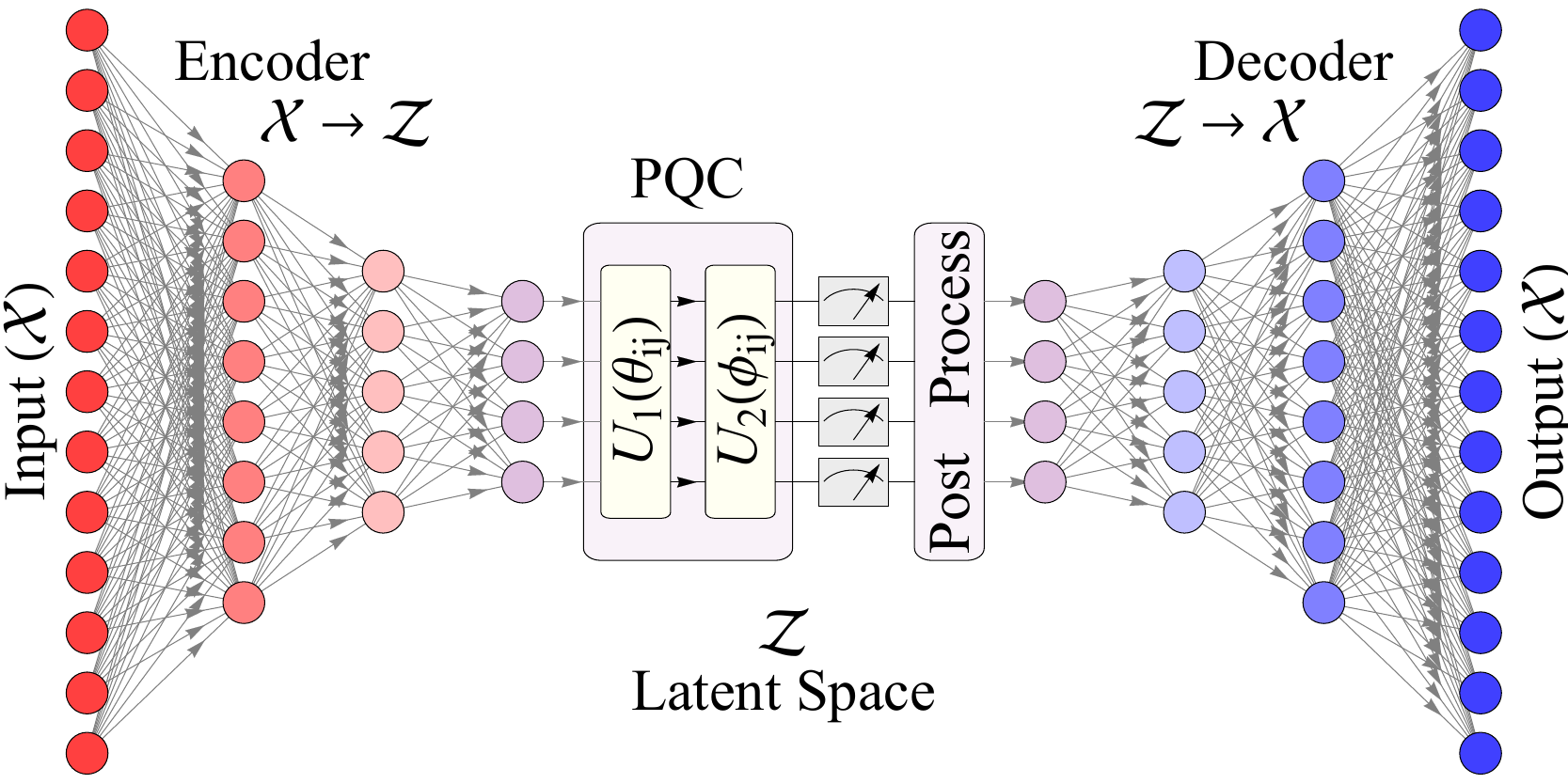}
\caption{A schematic diagram of a hybrid classical-quantum autoencoder (HAE) architecture \cite{sakhnenko2021hybrid}. The data coming from the classical encoder are embedded into the PQC. In PQC, the $U_1(\theta_{ij})$ and $ U_2(\phi_{ij})$ are the blocks of quantum circuits containing different rotation gates with rotation parameters $\theta, \phi$. After the blocks of quantum circuits there are measurements followed by the post measurement processing block (denoted as Post Process). After post-processing, the information is fed into the classical decoder.}
\label{fig:hybrid_autoencoder}
\end{figure}

The  quantum-classical hybrid anomaly detection scheme \cite{Romero_2017, qGAN4anomaly, anogan, sakhnenko2021hybrid} is the state of the art approach which utilises quantum machine learning \cite{Schuld2018, circuit_learning, transfer, funcke2021dimensional, Tabi2022} along with its classical counterpart. For our study we use the Hybrid Classical-Quantum Autoencoder (HAE) introduced by Sakhnenko et.al. in 2022 \cite{sakhnenko2021hybrid}, which significantly enhances the performance metrics of the anomaly detection compared to its fully classical counterpart. The HAE consists of a classical encoder, a parameterized quantum circuit (PQC)~\cite{pqc}, and a classical decoder (Fig.\,\ref{fig:hybrid_autoencoder}). The input goes to PQC via the encoder. The PQC is consists of quantum circuits containing different rotation gates. After the blocks of quantum circuits there are measurements followed by the post measurement processing block. After post-processing, the information is fed into the classical decoder.

It is worth mentioning that a PQC performs much better compared to an classical circuit with equal dimension \cite{sakhnenko2021hybrid}. Schuld \cite{hilbert_feature, schuld2021supervised} showed a connection between the quantum neural networks and kernel methods, where the quantum networks encode and process the data in a high-dimensional Hilbert space through a highly non linear feature mapping. This is classically intractable and only can be revealed through the inner products and measurements of the quantum states. In our case the PQC in HAE also expands the latent space into a higher dimensional Hilbert space. Therefore its internal degrees of freedom increases, resulting in a performance boost.

For HAE we consider the same encoder and decoder as the classical AE which is combined with a 4-qubit PQC (Fig.\,\ref{fig:hybrid_autoencoder}). The final quantum state is measured in the Pauli $Z$-basis, and the corresponding expectation value for each qubit construct the latent space for the anomaly detection. This information is fed to the decoder via the post processing module which expands the compressed data to its original size. The model is implemented using Qiskit \cite{Qiskit} for our analysis.

\subsection{Training and testing of IF, AE and HAE for the anomaly detection}\label{subsection:metrics_appendix}

For isolation forest (IF), we choose a training data set with nominal data points only. After training the model we apply the trained IF model on the testing data and obtain the predicted labels. For the AE and HAE, we first train the networks with the above training data for 50 epochs with batch size of 16 and learning rate 0.001, and then compute the losses during training.  We compute the mean squared error loss (MSE) defined as,
\begin{equation}
\text{MSE} = \frac{1}{n} \sum_{i=1}^n\left( x_i - \hat{x}_i \right)^2,
\end{equation}
where $x$ and $\hat{x}$ are the original and the predicted values of the datapoints respectively. By minimizing the above losses for the training data, we search for a suitable threshold with extensive empirical trials. Then we predict the outputs with respect to this threshold from the training data and compute the MSEs with respect to the actual training data. 
The threshold is then defined by the ratio between the mean and standard deviation of the MSEs. After defining the threshold we apply the trained model over the testing data and predict the outputs. We compute the MSEs with the predicted data and the actual test-data. For each prediction if the MSE is greater than the threshold then we label it as an anomaly/outlier, otherwise it is labelled as a nominal/normal data point. With the above training and testing procedure we finally compute the precision, recall, and F1 scores by comparing the predicted labels and the actual labels of the testing data.
 
\subsection{Performance measures of the model: precision-recall and F1 score}\label{subsection:Training_and_testing}

A reliable metric is necessary for measuring the performance of an anomaly-detection model, which should describe the fractions of uncovered anomalies from a mixture of nominal data and outliers. This is usually described by \textit{precision} (fraction of true anomalies of all discovered instances), \textit{recall} (fraction of true anomalies that were discovered), and their harmonic mean \textit{F1 score}  \cite{f_score, van1974foundation}, which are computed based on the counts of true positives ($TP$), false positives ($FP$), and false negatives ($FN$), are defined as follows:
\begin{eqnarray}
&precision = \frac{TP}{TP + FP}, ~ recall = \frac{TP}{TP + FN},& \nonumber \\
&F1~score = \frac{2 \times precision \times recall}{precision + recall}.&
\end{eqnarray}
An outcome with high \textit{recall} and low \textit{precision} contains more results but most of them would be wrong (FP).  A low \textit{recall} and high \textit{precision} on the other hand correspond less result but most of them would be right (TP). Most desirable outcome is one with both high \textit{recall} and high \textit{precision} which in turns give a high \textit{F1 score}.

\section{Results}\label{section:Results}

For our study we consider a two-terminal device configuration with randomly distributed 80 impurities (Fig.\,\ref{fig:1}). We use their coordinates ($x_1,y_1,x_2,y_2,\cdots$,  total 160 features) as the input and the resulting conductance ($T$) as the output. We consider two magnitudes of the impurity ($V_0$=0.3, 0.9).  Considering the distribution of conductance in these two cases (Fig.\,\ref{fig:1}), for $V_0$=0.9, $T>$0.5 is considered as anomaly, whereas for $V_0$=0.3, $T<$0.5 is considered as anomaly. 

For our study we consider two data sets, one for $V_0$=0.3 and one for $V_0$=0.9, each with 5000 different random configurations. In both cases total number of anomalies is less than 10\% of the entire data set. From each data set we prepare four different train-test samples by randomly choosing 900 nominal and 100 anomalous data points. After training and testing with four samples we compute the individual performance metrics (precision, recall, F1 score) and present their respective mean values 
in Table \ref{tab:summary_all_dataset}. For AE, input size is 160 (the original data size), which is followed by the encoder containing three layers with 106, 56, and 4 nodes respectively. The decoder is the mirror image of the encoder followed by the output layer of size 160 (Fig.\,\ref{fig:auto-encoder}). The latent size of our AE model is equal to 4 (Fig.\,\ref{fig:hybrid_autoencoder}). Same classical encoder and decoder is used for HAE with a 4-qubit PQC. A bigger latent dimension could improve the outcome, however due to the limitation of computational resources, we can't consider more than 4 qubit which also restricts the latent dimension of classical AE.

\begin{figure}[h!]
\centering
\includegraphics[width=\linewidth]{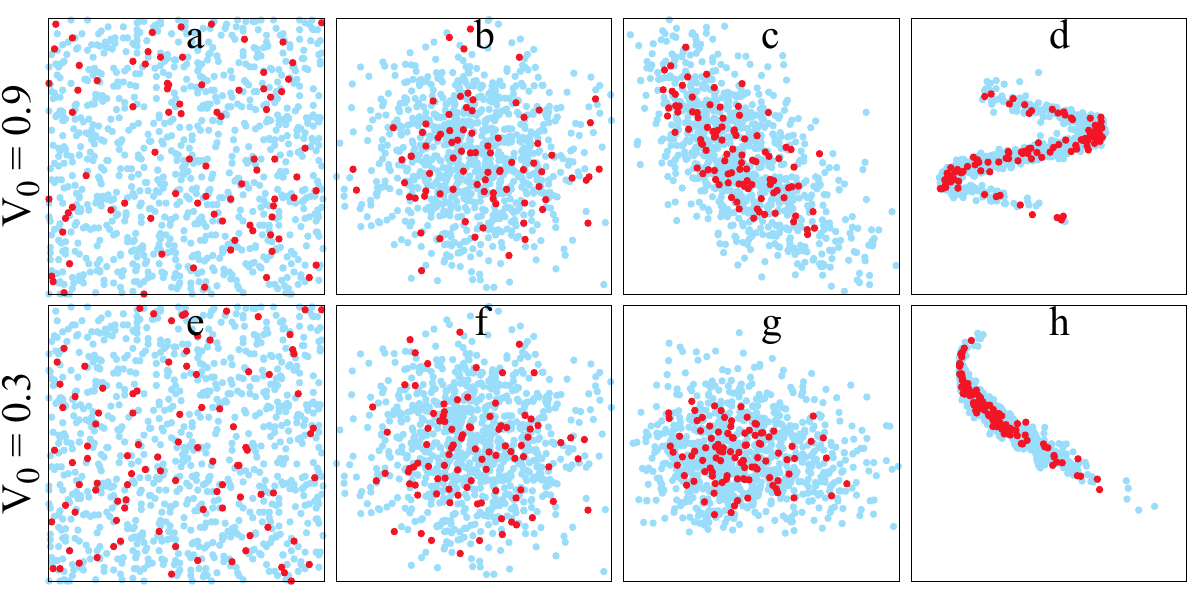}
\caption{Visualization of anomalies (red) and nominal data (light blue). Top (a,b,c,d) and bottom (e,f,g,h) panels present  $V_0$=0.9 and $V_0$=0.3 data sets respectively with respect to two arbitrary input dimension. (a,e) show the original input data. (b,f) show the PCA of the original data. (c,g) show the output from the classical encoder and (d,h) show the output from PQC.}
\label{fig:latent_space_cluster}
\end{figure}

Fig. \ref{fig:latent_space_cluster} shows two data sets with respect to two arbitrary features. Fig.\,\ref{fig:latent_space_cluster}a,e show the original data where the normal data (light blue) and anomalies (red) are uniformly distributed over the whole space. Fig.\,\ref{fig:latent_space_cluster}b,f show the output from PCA which show some clustering. This clustering is enhanced with a classical AE (Fig.\,\ref{fig:latent_space_cluster}c,g) and even more with a HAE (Fig.\,\ref{fig:latent_space_cluster}d,h). In a higher dimensional space such clustering leads to a better isolation of anomalies from the normal data which is also reflected in their individual performance metric (Table\,\ref{tab:summary_all_dataset}).


\begin{table}[h!]
\centering
\begin{tabular}{|c|c|c|c|c|}
\hline
\textbf{Set} & \textbf{Metric} & \textbf{IF}  & \textbf{AE}  & \textbf{HAE} \\ 
\hline
          & Precision  & 0.474 & 0.494	& 0.484 \\
$V_0$=0.9 & Recall	   & 0.675 & 0.715	& 0.748 \\
          & F1 score   & 0.556 & 0.583	& 0.588 \\
\hline
          & Precision  & 0.431 & 0.436	& 0.432 \\
$V_0$=0.3 & Recall	   & 0.497 & 0.595	& 0.618 \\
          & F1 score   & 0.461 & 0.503	& 0.509 \\
\hline
\end{tabular}
\caption{Performance metrics for anomaly detection IF, AE, and HAE using two data sets ($V_0$=0.3 and $V_0$=0.9). Each value shows the mean of the corresponding metric.
}
\label{tab:summary_all_dataset}
\end{table}
\noindent

Note that the data set corresponding to $V_0=0.9$ shows better performance compared to data set corresponding to $V_0=0.3$. This can be understood from the physical nature of the problem as well. For $V_0=0.9$ most of the configuration lead to a localization reducing the conductance. This is reflected in large occurrence of zero conductance in Fig.\,\ref{fig:1}a. Consequently the anomaly is very well defined since the majority of data has same characteristics. Compared to that, the distribution due to $V_0=0.3$ is relatively flat which makes the boundary between the nominal and anomalous data more smudged. Nevertheless, both AE and HAE show exceptional performance in both cases. To benchmark our results we compare them with the results obtained with other data sets. We find that the performance metrics obtained in Table \,\ref{tab:summary_all_dataset} are comparable to what is observed with standard publicly available data sets and even better than the dataset used to detect anomaly for gas turbine using same models \cite{sakhnenko2021hybrid}. From table \ref{tab:summary_all_dataset}, we see that the HAE is performing better in terms of recall and F1 scores keeping the precision  comparable to the other models. This is expected due to the inclusion of PQC as we discussed previously. Due the limitation of computational resource, we are limited to a 4 qubit PQC where the distinction between the performance of AE and HAE is not very prominent. By the increasing the dimension of the PQC and the latent space one can further enhance the performance of the HAE.

\section{Conclusion}

In this paper we demonstrate the application of anomaly detection to reveal exotic features of a condensed matter system. We consider an  Anderson model which shows two kind of anomalies, i.e. a high transmission at strong impurity strength and low conductance at weak impurity strength. Here we focus on three different approaches namely isolation forest (IF) which is based on the classification scheme random forest, auto encoder (AE) which is based on classical neural network and the hybrid classical-quantum auto encoder (HAE) which is a combination of a classical neural network and a parametric quantum circuit. Unlike classification scheme, here the training is done only on the normal data and the learning algorithm detect the anomalous outcome without any prior knowledge of the anomaly class. Performance of these algorithms are quantified with three different scores namely precision, recall, and F1 score. Predicting such high level of non-linear outcome is only possible via a neural network which is also reflected in their individual scores. We also demonstrate that the HAE performs better compared to its classical counterpart (AE) due to its inherent ability to deal with highly non linear feature mapping \cite{hilbert_feature, schuld2021supervised} which can't be achieved with classical circuit with same dimension.

In context of quantum transport, these anomaly detection schemes can be instrumental in understanding the behaviour of Anderson localisation as well as formation of solitons in disordered system. The method we present here is quite generic and can be extended to other systems. For example, in case of optical lattices, this formalism can be exploited to investigate Anderson localisation of light \cite{Segev2013}. For an abstract higher dimensional phase space, where the input parameters are made of different physical observable such as electronic or chemical properties of a system this approach can reveal new exotic configurations which can not be explored with any conventional methods, and thus would be instrumental in new physics hidden in remote corners of complex phase space.

\bibliographystyle{apsrev4-2}
\bibliography{ref}

\end{document}